\documentstyle[aps,prl,twocolumn]{revtex}

\newcommand{\CC}{{\cal C}}
\newcommand{\calB}{{\cal B}}
\newcommand{\la}{\langle}
\newcommand{\ra}{\rangle}
\newcommand{\be}{\begin{equation}}
\newcommand{\ee}{\end{equation}}
\newcommand{\ba}{\begin{array}}
\newcommand{\ea}{\end{array}}
\newcommand{\tr}{{\rm tr}}
\newcommand{\dd}{\mbox{Det}}
\newcommand{\hex}{{\rm H}}
\newcommand{\gr}{{\rm G}}
\newcommand{\Pp}{{\rm P}}
\newcommand{\Bb}{{\rm B}}
\newcommand{\Wt}{{\rm wt}}
\newcommand{\ghz}{{\rm GHZ}}

\begin{document}

\draft

\title{Entanglement entropy of multipartite pure states}

\author{Sergei Bravyi}

\address{e-mail:serg@itp.ac.ru \\
L.D. Landau Institute for Theoretical Physics\\
Kosygina St. 2, Moscow, 117940, Russia}

\date{\today}

\maketitle

\begin{abstract}
Consider a system consisting of $n$ $d$-dimensional quantum
particles and an arbitrary pure state $|\Psi\ra$ of the whole system.
Suppose we simultaneously perform complete von Neumann measurements
on each particle. One can ask: what is the minimal possible value
$S[\Psi]$ of the entropy of outcomes joint probability
distribution? We show that $S[\Psi]$ coincides with entanglement
entropy for bipartite states. We compute $S[\Psi]$ for two sample
multipartite states: the hexacode state $|{\rm H}\ra$\
($n=6$, $d=2$) and determinant states $|{\rm Det}_n\ra$ \ ($d=n$).
The result is $S[H]=4\log 2$ and $S[{\rm Det}_n]=\log(n!)$.
The generalization of determinant states to the case $d<n$
is considered.
\end{abstract}


\section{Introduction and main results}
\label{Intro}

Quantum information theory has many interesting features which have
no classical analogue. One of them is entanglement or quantum
correlations. It has been the object of intensive study for last
years because it is the entanglement that makes possible to develop
effective algorithms solving many tasks in computing, communication,
and cryptography, see for example~\cite{NC} and references therein.

However at the present moment the canonical definition of
entanglement is missing and the question how to quantify the degree
of entanglement in a given multipartite quantum state remains open.
The only reasonable constraint on a functional which pretends to be
an entanglement measure is the monotonicity under certain class of
local quantum operations~\cite{Vi}.

An important example of such functional is entanglement
entropy~\cite{BBPS}. For a pure state $|\Psi\ra$ of bipartite
system, its entanglement entropy $E[\Psi]$ is defined as
\be
\label{EE}
E[\Psi]=-\sum_i p_i \log_2 p_i,
\ee
where $p_i$'s denote the eigenvalues
of reduced density matrices $\rho_1 = \tr_2 (|\Psi\ra\la \Psi|)$
and $\rho_2 = \tr_1 (|\Psi\ra\la \Psi|)$ (they have the same spectrum).
This particular measure is distinguished because in the
asymptotic limit~(i.e. when one takes a large number of copies of a given
shared state) any monotonic functional of bipartite state up to trivial
rescaling coincides with entanglement entropy, see~\cite{PR}.
Unfortunately, when the system is
divided to three or more local parts,
entanglement entropy is not defined.

The functional $E[\Psi]$ has very simple physical sense. Suppose
that each of two parties, between which the state $|\Psi\ra$ is
distributed, performs complete von Neumann measurement on his part
of the system. Such joint measurement is a complete measurement on
the whole system. Its outcome is a random variable whose probability
distribution depends upon the pair of complete von Neumann
measurements chosen by each of two parties. This choice is
equivalent to the choice of orthonormal basis in each party
Hilbert space of states. One can ask: what bases should be chosen
by each party to minimize the entropy of the outcomes joint
probability distribution and what is the minimal value of this
entropy? If the state $|\Psi\ra$ is factorizable, i.e.
$|\Psi\ra=|\Psi_1\ra\otimes |\Psi_2\ra$, then the answer is
trivial: the $i$-th party should complement $|\Psi_i\ra$ to a
complete basis by any way and this choice yields zero entropy
because the measurement outcome will be "$(\Psi_1,\Psi_2)$" with
the probability one. If $|\Psi\ra$ is entangled, one can
easily show~(see Section~\ref{Bi}) that the $i$-th party should
perform measurement in the basis where its density matrix $\rho_i$
is diagonal~(the Schmidt basis) and the minimal entropy of the
outcomes coincides with $E[\Psi]$. Thus we can interpret
entanglement entropy $E[\Psi]$ as the minimum of outcomes entropy
over all choices of local complete von Neumann measurements.

If we consider entanglement entropy from this point of view,
it can be naturally  defined
for arbitrary multipartite states.
Suppose a system consists of $n$ $d$-dimensional
quantum particles distributed between $n$ remote
parties and let $|\Psi\ra \in (\CC^d)^{\otimes n}$
be arbitrary pure state of the whole system.
Denote $\calB_i$ the orthonormal basis in  $\CC^d$
chosen by $i$-th party and
let $|\calB_i (j)\ra \in \CC^d$ be the $j$-th basis vector in
the basis $\calB_i$,
where $i\in [1,n]$, $j\in[1,d]$.
Orthonormality condition implies that
$\la \calB_i (j) | \calB_i (j') \ra = \delta_{jj'}$.
Now let us define the functional $S[\Psi]$ according to:
\be
\label{S}
\ba{c}
S[\Psi] = \displaystyle \inf_{\calB_1, \ldots, \calB_n}
S[\Psi,\calB_1,\ldots,\calB_n],\\ \\
S[\Psi,\calB_1,\ldots,\calB_n] =
\displaystyle -\sum_j p(j)\log_2 p(j),\\  \\
\displaystyle
p(j)\equiv p(j_1,\ldots,j_n) =
|\la \Psi | \calB_1(j_1),\ldots, \calB_n(j_n) \ra|^2,\\
\ea
\ee
(the sum over multi-index $j$ is the sum over all
possible $j_1,\ldots,j_n \in [1,d]$).
It tells us to what extent  the parties may decrease
the entropy of the outcomes distribution by varying
the bases in which they perform the measurements.
As was said above, $S[\Psi]=E[\Psi]$ for bipartite states,
so we will call $S[\Psi]$ as entanglement entropy.
Its properties immediately following from
the definition are
$0\le S[\Psi] < n\log_2 d$; \
$S[\Psi]=0$ iff the state $|\Psi\ra$ is factorizable; \
$S[\Psi\otimes\Phi]=S[\Psi]+S[\Phi]$ (here we mean that
$|\Psi\ra$ and $|\Phi\ra$ are the states shared by $n$ and
$m$ parties, while $|\Psi\otimes\Phi\ra$ is shared by
$n+m$ parties); $S[\Psi]$ is continuous functional of the
state $|\Psi\ra$.

Computation of entanglement entropy in the multipartite case $n>2$
is a difficult task. Probably, for generic quantum state it can be
solved only numerically~(note that the number of parameters to be
optimized in the definition~(\ref{S}) grows as $O(nd^2)$). It is relatively easy
to get the upper bound on $S[\Psi]$ --- one just needs to choose
tentatively some basis for each party. As the lower bound on
$S[\Psi]$ one can take von Neumann entropy of the mixed state of
any group of parties~(see Section~\ref{Lower}). However for generic
state this lower bound is too weak. It can be improved if the state
has some symmetry.

 In this paper we consider two types of symmetry
on example of {\it determinant states} $|\dd_n\ra\in
(\CC^n)^{\otimes n}$, see~\cite{DW}, and six qubit {\it hexacode
state} $|\hex\ra$, see~\cite{CS}. The determinant state $|\dd_n\ra$
is invariant under unitary transformations $U^{\otimes n}$ where
$U\in \rm{SU}(n)$ is arbitrary one-party unitary operator. The
hexacode state is in some sense 'maximally uniform' pure state ---
if we divide six qubits into two equal groups by {\it arbitrary}
way then the mixed state of each group will be absolutely uniform.
Due to these special properties, entanglement
entropy of the states $|\dd_n\ra$ and $|\hex\ra$ can be exactly
computed.

Another reason for our interest to these particular states is that
their entanglement entropy is rather close to the upper bound
$n\log_2 d$, so they are near-maximally entangled states. For the
hexacode state $(n=6,\, d=2)$ the computation yields $S[H]=4$,
see Section~\ref{Hex}, while for determinant state $(n=d)$ one
gets $S[\dd_n]=\log_2 (n!)$, see Section~\ref{Det}. For large $n$ we
can approximately write $S[\dd_n]\approx [1-1/\ln(n) ]\, n\log_2
n$. It means that determinant states asymptotically saturate the
upper bound for normalized entropy: $\lim_{n\to \infty}
S[\dd_n]/n\log_2 n =1$.
In
Section~\ref{Gdet} we construct the generalized determinant
states $|\dd_{n,d} \ra \in (\CC^d)^{\otimes n}$ defined if
$n=pd^p$, where $p$ is arbitrary integer, such that
$S[\dd_{n,d}]=\log_2[ (d^p)! ]$. For fixed $d$ and large $n$~(i.e.
large $p$) we can write: $S[\dd_{n,d}]\approx
[1-1/\ln(n) ]\, n\log_2 d$ which again saturates the upper bound
for normalized entropy. Note that the factor $[1-1/\ln(n)]$ grows
sufficiently slow, e.g. it is equal 0.9 for $n=e^{10}\approx 2\cdot
10^4$. We will see that for qubits~($d=2$) the tensor powers of
hexacode state $|\hex^{\otimes m}\ra$ have entanglement entropy
greater than determinant states $|\dd_{n,2}\ra$ if the number of
qubits $n \lesssim 60$.

We make concluding remarks and discuss some open questions
concerning the generalized entanglement entropy in
Section~\ref{Concl}. Most interesting question concerns the
stability of the definition~(\ref{S}) under the extension of each
party space of states.

\section{Bipartite system}
\label{Bi}

It is known that up to local
unitary operators, any state
$|\Psi\ra \in \CC^d \otimes \CC^d$ of
bipartite system is specified by its Schmidt coefficients
$\{ p_i \}_{i=1\ldots d}$, $p_i \ge 0$, $\sum_{i=1}^d p_i =1$.
Being invariant under local unitaries
$S[\Psi]$ is a functional of the Schmidt coefficients only.
Thus one suffices to compute $S[\Psi]$ only for the special states
\be
|\Psi\ra = \sum_{i=1}^d \sqrt{p_i} \, |i,i\ra,
\ee
where $\{ | i\ra \in \CC^d \}_{i=1\ldots d}$ is
the standard basis of $\CC^d$.
If in the definition~(\ref{S}) we tentatively choose the standard
basis for both parties, i.e.
$|\calB_1 (i)\ra = |\calB_2 (i)\ra = |i\ra$,
$i\in [1,d]$, then
$S[\Psi,\calB_1,\calB_2] = -\sum_{i=1}^d p_i \log_2 p_i$
and thus we get
$S[\Psi] \le E[\Psi]$,
see~(\ref{EE}).
We can also prove
that $E[\Psi]$ is simultaneously the lower bound for $S[\Psi]$.
Indeed, let $\calB_1^*$, $\calB_2^*$ be the optimal choice
of the bases
(i.e. such that $S[\Psi]=S[\Psi,\calB_1^*,\calB_2^*]$).
Consider the density matrix $\rho_1$ of the first party only:
$\rho_1= \tr_2(|\Psi\ra\la\Psi|)=\sum_{i=1}^d p_i |i\ra\la i|$.
Denote $p_1(i) = \la \calB_1^*(i)| \rho_1 | \calB_1^* (i)\ra
= \sum_{j=1}^d p_j |\la \calB_1^* (i) |j\ra|^2$
the distribution of the first party outcomes
in the optimal basis.
Because the entropy of partial distribution
can not exceed the entropy of joint distribution,
we have:
\be
\label{ineq1}
S[\Psi] \ge - \sum_{i=1}^d p_1(i) \log_2 [p_1(i)].
\ee
Using the concavity of the function $-x\log_2 x$ and
normalization $\sum_{j=1}^d |\la \calB_1^* (i) |j\ra|^2 =1$,
we get the next estimate:
\be
\label{ineq2}
 - p_1(i) \log_2 p_1(i) \ge -
\sum_{j=1}^d |\la \calB_1^* (i)|j\ra|^2 (p_j \log_2 p_j ).
\ee
The summation over $i$ can be carried out
taking into account the normalization
$\sum_{i=1}^d |\la \calB_1^* (i) |j\ra|^2 =1$.
Thus from~(\ref{ineq1},\ref{ineq2}) we can infer
$S[\Psi]\ge E[\Psi]$
and consequently
\be
\label{bi}
S[\Psi]=E[\Psi].
\ee

\section{Connection with von Neumann entropy}
\label{Lower}

Suppose the system consists of $n$ $d$-dimensional
quantum particles distributed between $n$ remote
parties and let $|\Psi\ra \in (\CC^d)^{\otimes n}$
be arbitrary pure state of the whole system.
Let us choose a group $X$
of $k$ parties, for example $X=\{1,2,\ldots,k\}$.
The chosen group of parties shares the mixed state
$\rho_x = \tr_{j\notin X} (|\Psi\ra\la\Psi|)$.
Suppose $\calB^*_i$ is the optimal basis for the $i$-th party.
Denote $p_x(i_1,\ldots,i_k)\equiv p_x(i)$
the optimal outcomes distribution for the parties from $X$ only:
\be
\label{p_X}
p_x(i)=\la \calB^*_1(i_1),\ldots,\calB^*_k(i_k)|
\rho_x |\calB^*_1(i_1),\ldots,\calB^*_k(i_k)\ra.
\ee
Let $S[p_x(i)]$ be the entropy of distribution~(\ref{p_X}).
By repeating the arguments presented in Section~\ref{Bi} we can
show that
\be
\label{low}
-\tr( \rho_x \log_2 \rho_x) \le S[p_x(i)] \le S[\Psi].
\ee
Thus von Neumann entropy of the mixed state $\rho_x$ can serve
as the lower bound on entanglement entropy. Of course the
group of parties $X$ can be chosen by arbitrary way.

Note that the density matrix of the parties which were not
selected to $X$ has the same~(positive) spectrum as $\rho_x$.
It means that one suffices to consider the groups of
$k\le n/2$ parties and the best lower estimate on $S[\Psi]$
which we can hope to achieve is $(n/2) \log_2 d$.

As a good  example, consider three qubit
GHZ state $|\ghz\ra =2^{-\frac12} (|0,0,0\ra+
|1,1,1\ra)$, $d=2$, $n=3$. The density matrix of the first
qubit is absolutely uniform: $\rho_1 = (1/2)\hat{1}$.
Thus $S[\ghz]\ge -\tr(\rho_1 \log_2 \rho_1) =1$.
On the other hand we can choose tentative bases
$\calB_1=\calB_2$ $=\calB_3=\{|0\ra,|1\ra\}$ which
provide us with
upper estimate $S[\ghz]\le
S[\ghz,\calB_1,\calB_2,\calB_2]=1$, so that
$S[\ghz]=1$.

\section{Determinant state}
\label{Det}

Let us consider the multipartite system with $d=n$.
Choose the standard
basis
$\{ |i\ra \in \CC^n \}_{ i=1\ldots n}$ in each copy of
$\CC^n$ and consider the state $|\dd_n\ra \in (\CC^n)^{\otimes n}$
defined as
\be
\label{det_def}
|\dd_n\ra = (n!)^{-\frac12} \sum_{i_1,\ldots,i_n }
\epsilon_{i_1,\ldots,i_n} |i_1,\ldots,i_n\ra,
\ee
where $\epsilon_{i_1,\ldots,i_n}$ is completely antisymmetric
tensor of the rank $n$ and the sum is over all $i_1,\ldots,i_n \in
[1,n]$. This state was used in Ref.~\cite{DW} to study the
limitations on the pairwise entanglement in multipartite systems.
We will call the family $\{ |\dd_n\ra \}_n$ the determinant states.
Note that $|\dd_2\ra$ is EPR singlet state
$2^{-\frac12}(|1,2\ra-|2,1\ra)$. The purpose of this section is to
prove the formula
\be
\label{factor1}
S[\dd_n]=\log_2 (n!).
\ee
This result is immediate consequence of
the following property of determinant states:
\be
\label{aux}
\sup_{
\ba{c}
\scriptstyle |\phi_i\ra \in \CC^n,\\
\scriptstyle \la \phi_i|\phi_i\ra=1,\, \, i\in [1,n].\\ \ea }
|\la \dd_n |\phi_1,\ldots,\phi_n \ra|^2=(n!)^{-1},
\ee
(we employ standard designation
$|\phi_1,\ldots,\phi_n \ra\equiv
|\phi_1\ra$ $\otimes\cdots\otimes|\phi_n\ra$).
It tells us that the projection of $|\dd_n\ra$ on any factorizable
state has the norm  at most $(n!)^{-1}$.
To prove~(\ref{aux}), we first
note that the state $|\dd_n\ra$ is $\mbox{SU}(n)$ singlet, i.e. for
any $V\in \mbox{SU}(n)$ we have $V^{\otimes n} |\dd_n\ra$
$=|\dd_n\ra$. Therefore, if the projection of $|\dd_n\ra$ on the
state $|\phi_1,\ldots,\phi_n\ra$ is the highest one,
then the projection
of $|\dd_n\ra$ on the state $|V\phi_1,\ldots, V\phi_n\ra$
is also the highest one. So while looking for the
maximum in~(\ref{aux}) we can fix one of $|\phi_i\ra$, e.g. put
$|\phi_n\ra=|n\ra$. But according to definition~(\ref{det_def}) we
have:
\be
\label{induct}
|\la \dd_n |\phi_1,\ldots,\phi_{n-1},n\ra|^2 =
\frac1{n} |\la \dd_{n-1} |\phi_1,\ldots,\phi_{n-1}\ra|^2.
\ee
Here by abuse of notations we denote $|\dd_{n-1}\ra$ the embedding
of determinant state $|\dd_{n-1}\ra\in (\CC^{n-1})^{\otimes(n-1)}$
into the space $(\CC^n)^{\otimes(n-1)}$~(the space $\CC^{n-1}$
is embedded into $\CC^n$ by adding zero $n$-th
component to all vectors).
Although in~(\ref{induct}) $|\phi_i\ra \in \CC^n$,
the righthand side of~(\ref{induct}) achieves the
maximum when all states $|\phi_1\ra,\ldots,|\phi_{n-1}\ra$ have
zero $n$-th component.
It implies that
$$
\ba{rcl}
& \displaystyle
\quad \sup_{
\ba{c}
\scriptstyle |\phi_i\ra \in \CC^n,\\
\scriptstyle \la \phi_i|\phi_i\ra=1,\, \, i\in [1,n].\\ \ea } &
|\la \dd_n |\phi_1,\ldots,\phi_n \ra|^2
\\ \\
{} = & \displaystyle \frac1{n}
\sup_{
\ba{c}
\scriptstyle |\phi_i\ra \in \CC^{n-1},\\
\scriptstyle \la \phi_i|\phi_i\ra=1,\, \, i\in [1,n-1].\\ \ea }&
|\la \dd_{n-1} |\phi_1,\ldots,\phi_{n-1} \ra|^2,
\\
\ea
$$
which by induction leads to equality~(\ref{aux}).

Now let us explain why~(\ref{aux}) implies~(\ref{factor1}). Suppose
$\calB_i^*$ is the optimal basis for the $i$-th copy of $\CC^n$,
$i\in [1,n]$, i.e.
$S[\dd_n]=S[\dd_n,\calB_1^*,\ldots,\calB_n^* ]$. Let
$p^*(i_1,\ldots,i_n)$ $=|\la \calB_1^* (i_1),\ldots,\calB_n^*
(i_n)|\dd_n\ra|^2$ be the optimal distribution. According
to~(\ref{aux}), $p^*(i_1,\ldots,i_n)\le (n!)^{-1}$ for any outcomes
$i_1,\ldots,i_n$ and thus $S[\dd_n]\ge \log_2 (n!)$.
On the other hand, we can tentatively
suggest all parties to perform the measurements in the standard basis,
i.e.
$|\calB_i (j)\ra = |j\ra$, $i,j \in [1,n]$.
Then $S[\dd_n,\calB_1,\ldots,\calB_n]$ $=
\log_2(n!)$ which tells us that $S[\dd_n]\le \log_2 (n!)$
and thus that $S[\dd_n]=\log_2 (n!)$.

\section{Generalized determinant state}
\label{Gdet}

Suppose now that the dimension $d$ of each particle is fixed. If the
number of particles is $n=pd^p$ for some integer $p$, the
space $(\CC^d)^{\otimes n}$ can be identified with
$(\CC^{d^p})^{\otimes d^p}$ and thus determinant state
$|\dd_{d^p}\ra$ has its counterpart in $(\CC^d)^{\otimes n}$. This
simple observation allows to construct the state $|\dd_{n,d}\ra \in
(\CC^d)^{\otimes n}$, $n=pd^p$ such that
\be
\label{factor2}
S[\dd_{n,d}] = \log_2 [ (d^p)! ], \ \ n=pd^p.
\ee
Note that although $|\dd_{d^p}\ra$ and $|\dd_{n,d}\ra$
represent one and the same state, $S[\dd_{n,d}]$ might be
greater than $S[\dd_{d^p}]$ because in the first case we have less
freedom in the choice of bases in~(\ref{S}).

Let us explain the construction of the state $|\dd_{n,d}\ra$
and derive~(\ref{factor2}) on
example of the qubits, i.e. $d=2$. Consider any one-to-one map
$\varphi$ which maps the integers on the interval $[1,2^p]$ to
binary strings of the length $p$, e.g.
\be
\label{varphi}
\ba{rcl}
\varphi(1) & = & (0,0,\ldots,0,0), \\
\varphi(2) & = & (0,0,\ldots,0,1), \\
& \cdots & \\
\varphi(2^p-1) & = & (1,1,\ldots,1,0), \\
\varphi(2^p) & = & (1,1,\ldots,1,1). \\
\ea
\ee
Define the state $|\dd_{n,2}\ra \in (\CC^2)^{\otimes n}$, $n=p2^p$ as
\be
\label{det_def1}
|\dd_{n,2}\ra \sim \sum_{i_1,\ldots,
i_{2^p}}
\epsilon_{i_1,\ldots,i_{2^p}}
|\varphi(i_1),\ldots,\varphi(i_{2^p})\ra,
\ee
where the sum is over all $i_1,\ldots,i_{2^p} \in [1,2^p]$
and we omit the normalizing factor $[(2^p)!]^{-\frac12}$.
It is written here in the standard qubit basis $\{ |0\ra, |1\ra \}$.
While computing $S[\dd_{n,2}]$ we
minimize over the choice of $p2^p$ one-qubit bases, see~(\ref{S}).
But any such choice is also the choice of $2^p$ bases in $2^p$
copies of $\CC^{2^p}$. Therefore we can say that
\be
\label{great}
S[\dd_{n,2}]\ge S[\dd_{2^p}]=\log_2 [ (2^p)! ].
\ee
On the other hand, if we will tentatively measure each of $p2^p$
qubits in the basis $\{ |0\ra, |1\ra\}$, the entropy of the
outcomes distribution will be exactly $\log_2 [ (2^p)! ]$,
see~(\ref{det_def1}). Therefore $S[\dd_{n,2}]\le \log_2 [ (2^p)! ]$
and thus the equality~(\ref{factor2}) is proven~(the proof for
arbitrary $d$ copies the proof for $d=2$).

As was already mentioned in Section~\ref{Intro}, the determinant
states are near-maximally entangled states if the number of
parties $n$ is sufficiently large. As an illustration let us consider
determinant states $|\dd_{n,2}\ra$ corresponding to
$n$ qubit system.
In Table~I we present the normalized entanglement entropy
$S[\dd_{n,2}]/n$ and the number of qubits $n$ for $p=1,\ldots,5$,
and $p=10$.
\begin{table}
\begin{center}
\caption{Normalized entanglement entropy of determinant
states $|\dd_{n,2}\ra$. Here $n=p2^p$ is the number of qubits.}
\end{center}
\begin{center}
\begin{tabular}{ccccc}
& $p$ & $n$ & $S[\dd_{n,2}]/n$ & \\
\hline
& $1$ & $2$ & $0.50$ & \\
& $2$ & $8$ & $0.57$ & \\
& $3$ & $24$ & $0.64$ & \\
& $4$ & $64$ & $0.69$ & \\
& $5$ & $160$ & $0.74$ & \\
& $10$ & $10240$ & $0.86$ & \\
\end{tabular}
\end{center}
\end{table}
\noindent
Table~I suggests that we could try to find some state of $O(1)$
qubits which has entanglement entropy greater than the determinant
states. This is the purpose of the next section.

\section{Hexacode state}
\label{Hex}

The hexacode state was originally defined in the context of quantum
error correcting codes. It was associated with certain maximal
self-dual linear subspace of $GF(2)^6$, see~\cite{CS}, p.30 for
details. In this Section we present the alternative and more
explicit definition of this state only briefly discussing its
connection
with quantum codes. We also prove the equality
\be
\label{H}
S[\hex]=4,
\ee
announced in Section~\ref{Intro}.

Consider graph $G=(V,E)$ shown on Fig.~1 with the
set of vertices $V=\{1,2,3,4,5,6\}$
and the set of edges $E=\{(12),(13),(14),\ldots,
(56)\}$.
\begin{center}
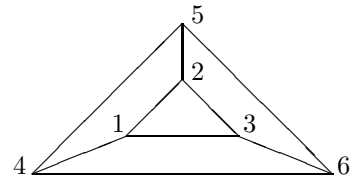
\begin{figure}
\unitlength=2.5mm
\begin{picture}(16,10)(0,-2)
\put(0,0){\line(1,0){16}}
\put(0,0){\line(1,1){8}}
\put(8,8){\line(1,-1){8}}
\put(0,0){\line(5,2){5}}
\put(16,0){\line(-5,2){5}}
\put(8,8){\line(0,-1){3}}
\put(5,2){\line(1,0){6}}
\put(5,2){\line(1,1){3}}
\put(8,5){\line(1,-1){3}}
\put(-1,0){$4$}
\put(16,0){$\,6$}
\put(8,8){$\ 5$}
\put(4,2.2){$\,1$}
\put(11,2.2){$\,3$}
\put(8,5){$\ 2$}
\end{picture}
\caption{Graph $G$ used in the definition of the hexacode state.}
\end{figure}
\end{center}
We associate a qubit with each vertex $i\in V$.
Let $A_{ij}$ be 6x6
adjacency matrix of $G$, i.e. $A_{ij}=1$ if $(ij)\in E$ and
$A_{ij}=0$ if $(ij)\notin E$, $A_{ij} =A_{ji}$.
The diagonal elements $A_{ii}$ will not appear anywhere.
Then six qubit hexacode state $|\hex\ra
\in (\CC^2)^{\otimes 6}$
is defined as follows:
\be
\label{hex_def}
\ba{c}
\displaystyle
|\hex\ra =
(2^6)^{-\frac12}
\sum_{x\in B^6}
(-1)^{a(x)}\,
|x\ra,
\\ \\
\displaystyle
a(x)=\sum_{i<j} A_{ij} x_i x_j,\\
\ea
\ee
where $B^6$ denotes the set of all binary strings of the length 6
and $|x\ra=|x_1,\ldots,x_6\ra$.

Note that $|\hex\ra$ can also be defined in terms of stabilizers
operators. Denote $\sigma^{\alpha}_i$ the Pauli matrix
$\sigma^\alpha$ acting on the $i$-th qubit and assign to each
vertex of the graph the operator
\be
\label{stabilizer}
X_i = \sigma^x_i \prod_{j \, : \, (ij)\in E}
\sigma^z_j,  \ \ i\in V.
\ee
They commute with each other and stabilize the state $|\hex\ra$,
i.e. $X_i |\hex\ra = |\hex\ra$. Six operators $X_i$ generate the
group $S$ of all stabilizers, $|S|=2^6$. Each stabilizer from $S$
is a tensor product of several Pauli matrices (probably with '-' sign).
As was shown in~\cite{CS}, any nontrivial stabilizer from $S$ is tensor product
of at least four Pauli matrices, so that $|\hex\ra$
is an additive
quantum code coding 0 qubits into 6 qubits with the minimal stabilizer
weight $4$. Of course, such quantum code
can not be used to
protect quantum information from the errors. It is just
symplectic state with some special properties. Note however
that if some  symplectic state has the minimal stabilizer weight $d$
and if any $[(d-1)/2]$ or less qubits were decohered then the
syndrom measurement allows to determine the positions of
the decohered qubits.

The proof of the equality~(\ref{H}) consists of three steps.
The first step is to prove that $S[\hex]\le 4$.
On the second step we establish remarkable symmetry of the state
$|\hex\ra$ which is used on the third step
to prove that $S[\hex]\ge 4$.

1) Let us tentatively choose the bases $\calB_i = \{ |0\ra, |1\ra
\}$ for the qubits $i=2,3,4,5$ and $\calB_1=\calB_6=\{ |+\ra, |-\ra
\}$ where $|\pm\ra = 2^{-\frac12} (|0\ra \pm |1\ra)$. Simple
calculations show that the distribution of the outcomes $p(x)\equiv
p(x_1,\ldots,x_6)$ $=|\la \calB_1(x_1),\ldots,\calB_6(x_6)|\hex\ra|^2$
measured in these bases is following:
\be
\label{best}
p(x)= \left\{
\ba{ll}
\frac{1}{16}, & \mbox{if}\
\left\{
\ba{l}
x_1=x_2+x_3+x_4\ (mod\, 2), \\
x_6 = x_3 +x_4+x_5\ (mod\, 2),\\
\ea \right.\\ \\
0, & \mbox{otherwise}.\\
\ea \right.
\ee
Thus $S[\hex,\calB_1,\ldots,\calB_6]=4$ and consequently
$S[\hex]\le 4$.

2) Suppose the vertices of $G$ are colored by black and white
colors such that there are three black and three white vertices.
Denote $B$ and $W$ the subsets of black and white vertices, $B\cup
W = V$. Consider three-qubit density matrix $\rho_w$ describing the
state of white qubits only: $\rho_w = \tr_B (|\hex\ra \la \hex |
)$. We claim that for
 any
partition $V=B\cup W$, $\rho_w = 2^{-3}\, \hat{1}$ where $\hat{1}$
is the unital matrix. In other words, the mixed state of any triple
of the qubits is absolutely uniform. To verify this property, fix
some partition  and renumber the vertices of the graph to make
$W=\{1,2,3\}$ and $B=\{4,5,6\}$. The adjacency matrix $A$ then can
be split to four 3x3 blocks:
\be
\label{blocks}
A=\left(
\ba{lr}
A^{ww} & A^{wb} \\
A^{bw} & A^{bb} \\
\ea
\right)
\ee
which are adjacency matrices between white and white, white and
black, black and white, black and black vertices
($A^{wb}=(A^{bw})^T$). The density matrix $\rho_w$ depends only
upon $A^{ww}$ and $A^{bw}$. Simple calculations yield:
\be
\label{rho_w}
\ba{rcl}
\la x| \rho_w |y\ra  & = &
(-1)^{ x^T A^{ww} x\, + \, y^T A^{ww} y}\\
&& {} \displaystyle \cdot 2^{-6}
\sum_{z\in \Bb^3}
(-1)^{z^T A^{bw} (x+y)}.\\
\ea
\ee
where $x\equiv(x_1,x_2,x_3)$, $y\equiv(y_1,y_2,y_3)$,
$z\equiv(z_1,z_2,z_3)$ and we treat $A^{ww}$, $A^{bw}$ as 3x3
matrices over binary field acting on binary vectors. Observe that
the sum over $z$ is zero if $A^{bw}(x+y)\ne (0,0,0)$ and is equal
to $2^3$ if $A^{bw}(x+y)=(0,0,0)$. One can explicitly verify that
the matrix $A^{bw}$ is nondegenerate over binary field~\cite{bin}
for any partition $V=B\cup W$~
(note that due to the symmetry
of the graph $G$, there are only three nonequivalent partitions,
so this verification is very simple).
It means that $A^{bw}(x+y)=(0,0,0)$ only if $x=y$.
Therefore $\la x
|\rho_w |y\ra =0$ if $x\ne y$ and $\la x|\rho_w |x\ra = 2^{-3}$,
i.e. $\rho_w = 2^{-3}\, \hat{1}$, see also~\cite{est}.

3) Suppose the optimal basis for the $i$-th qubit is $\calB^*_i$,
i.e. $S[\hex]=S[\hex,\calB^*_1,\ldots,\calB^*_6]$. Let
$p^*(x_1,\ldots,x_6)$ $=|\la
\calB_1^*(x_1),\ldots,\calB_6^*(x_6)|\hex\ra|^2$ be the optimal
probability distribution. Consider any partition $V=B\cup W$. The
probability distribution of three outcomes
$ \{ x_i \, : \,  i\in W\}$
measured at white vertices only is
\be
\ba{l}
\displaystyle
\sum_{x_i \, : \,  i\in B} p^*(x_1,\ldots,x_6) \\
\displaystyle
\qquad \quad {} = \la
\otimes_{i\in W}
\calB^*_i(x_i) |
\rho_w
|
\otimes_{i\in W}
\calB^*_i(x_i)
\ra
= \frac18, \\
\ea
\ee
regardless of configuration $\{ x_i \, : \, i\in W\}$.
In other words, the optimal distribution has a nice property:
the partial distribution of any triple of bits is absolutely
uniform, see~\cite{remark}.
 Call such property of the distribution as
3-uniformity. An example of 3-uniform distribution is absolutely
uniform distribution. There are also 3-uniform distributions which
are not absolutely uniform, e.g. the distribution~(\ref{best}).
 Denote $P_6^3$ the set of all
3-uniform distributions of six bits. In Appendix~\ref{App} we show that
\be
\label{inf6}
\inf_{p(x) \in P_6^3}
S[p(x)] =4,
\ee
where $S[p(x)]$ is the entropy of probability distribution $p(x)$.
We know that $p^*(x_1,\ldots,x_6) \in P_6^3$ and thus we have
$S[\hex]= S[p^*(x_1,\ldots,x_6)]\ge 4$. This completes the proof
of~(\ref{H}).

The definition like~(\ref{hex_def}) can be used to assign a state
$|\gr\ra\in (\CC^2)^{\otimes |V|}$ to any unoriented graph
$G=(V,E)$. Reasoning as above, one can show that if $|V|=2m$ and
the adjacency matrix $A^{bw}$ is nondegenerate over binary field
for any partition $V=B\cup W$, $|B|=|W|=m$, then in the state
$|\gr\ra$ any $m$ qubits have absolutely uniform density matrix
$2^{-m}\, \hat{1}$. Such state $|\gr\ra$ can be called maximally
uniform pure state, because any subset of qubits which is not
forbidden by Schmidt constraint to have absolutely uniform density
matrix do have absolutely uniform density matrix.
 This property of the quantum
state is interesting by itself. Surprisingly, for $m\le 15$ (i.e.
for $|V|\le 30$) appropriate graphs exist only for $m=3$ (e.g. the
graph shown on Fig.~1) and for $m=1$ (e.g. $V=\{1,2\}$ and
$E=\{(12)\}$). It follows from the bounds on additive quantum
codes.
 Indeed, consider a state $|\gr\ra$ assigned
to such graph. Like as hexacode state, we can specify $|\gr\ra$ by
stabilizer  operators~(\ref{stabilizer})
which generate the group $S$ of stabilizers of order $2^{2m}$.
Any stabilizer $X\in S$ is a tensor product of several Pauli matrices
(possibly with a sign '-') and
 $X|\gr\ra=|\gr\ra$.
But for any operator $Y$ acting on $m$ or less qubits
we have $\la \gr|Y|\gr\ra=2^{-m}\tr(Y)$
because the density matrix of any $m$ qubits is proportional
to $\hat{1}$. Thus any stabilizer
$X\in S$ acts on at least $m+1$ qubit.
It means that $|\gr\ra$ is an additive quantum code
coding 0 qubits into $2m$ qubits with the
minimal stabilizer weight $m+1$ or greater.
The results of the work~\cite{CS}
imply that for $m\le 15$ such codes exist only for $m=1,3$.

The tensor powers of hexacode state $|\hex^{\otimes k}\ra$
have the highest~(to our knowledge) entanglement entropy for
sufficiently small $k$. For example, the state $|\hex^{\otimes 4}\ra$
has entanglement entropy greater than determinant state
$|\dd_{24,2}\ra$, if  one takes twenty-four qubits, see Table~I.
However we do not expect that $|\hex\ra$ has the  maximal
entanglement entropy if all six qubit states could be
considered.

\section{Conclusion}
\label{Concl}

Although the generalization of entanglement entropy
to multipartite case suggested in the present work
looks rather natural, one faces a lot of difficulties
while trying to compute entanglement
entropy of some particular state.
A progress can be achieved only if the state has some
special properties or symmetry.

For fixed $n$ and $d$
the maximal  entanglement entropy
$S^*(n,d)$ is rather close to the upper bound
$n\log_2 d$, such that
the ratio $S^*(n,d)/n\log_2 d$ approaches to one
for fixed $d$ and sufficiently large $n$.

There is also one subtle point
in the definition~(\ref{S})
concerning its stability under the
extension of each party space of
states.
Suppose that each of the parties sharing the
state $|\Psi\ra\in (\CC^d)^{\otimes n}$
adds new local degrees of freedom
to his part of the system
thus extending his space to $\CC^D$,
$D>d$. The original space $\CC^d$ is
somehow embedded to extended space
$\CC^D$ and the original state $|\Psi\ra$
now is a vector from $(\CC^D)^{\otimes n}$.
There are two ways to compute entanglement
entropy of $|\Psi\ra$: the parties may
performe complete von Neumann measurements
either in the original space $\CC^d$ or
in the extended space $\CC^D$.
Being the functional of the state $|\Psi\ra$
only, entanglement entropy should be the
same in both cases.
If  $S[\Psi]$ is indeed invariant
under such extensions, call $|\Psi\ra$ the
stable state. In the bipartite case any state
is stable, because $S[\Psi]$ is invariant
functional of single party density matrix,
see~(\ref{EE},\ref{bi}).
The determinant state $|\dd_n\ra$ is also
stable for any $n$. Indeed, the formula~(\ref{aux})
which guarantees the equality $S[\dd_n]=\log_2 (n!)$
remains valid even if we allow the states
$|\phi_i\ra$ to be chosen from the extended space:
the maximum is obviously achieved when
all $|\phi_i\ra$'s belong to original space
$\CC^n$. However we can not prove that
arbitrary multipartite state is stable, so
this question is open.

Also it is interesting to check whether generalized
entanglement entropy is monotonic under
local quantum operations for $n>2$.

\begin{acknowledgements}
I would like to acknowledge fruitful discussions with Alexei
Kitaev, John Preskill, and David DiVincenzo during my visit to
Institute of Quantum Information, Caltech. I would also like to
thank Guifre Vidal for his contribution to investigation of the
properties of determinant states. Financial support from NWO-Russia
collaboration program is also aknowledged.
\end{acknowledgements}

\appendix
\section{}
\label{App}

The purpose of this section is to prove the equality~(\ref{inf6}).
We will consider probability distributions $p(x)$ of $n$ classical
bits, i.e. $x=(x_1,\ldots,x_n) $, $x_i=0,1$. By analogy with the
set $P_6^3$ we will consider the sets $\Pp_n^k$ of $k$-uniform
distributions of $n$ bits. By definition, $p(x)\in P_n^k$ iff any
$k$ of $n$ bits have absolutely uniform distribution~(i.e. if the
sum of $p(x)$ over any $n-k$ bits is equal to $2^{-k}$). For
example, the set $\Pp_n^n$ consists of just one point --- absolutely
uniform distribution of $n$ bits $p(x)\equiv 2^{-n}$ while the set
$\Pp_n^0$ includes all possible $n$-bit distributions. If $p(x)\in
\Pp_n^k$, consider its binary Fourier transform $q(y)$:
\be
\label{Fourier}
\left\{
\ba{l}
q(y)=\sum_{x\in \Bb^n} (-1)^{x\cdot y }\,
p(x), \\ \\
p(x)=2^{-n} \sum_{y\in \Bb^n}
(-1)^{x\cdot y}\,
q(y).\\
\ea
\right.
\ee
Here $\Bb^n$ denotes the set of all length $n$ binary strings and
$x\cdot y = \sum_{i=1}^n x_i y_i \, (mod\, 2)$.
If $y\in \Bb^n$, denote $\Wt(y)\in [0,n]$ the
number of '1' in the binary string $y$.
The definition of $k$-uniformity
can be rephrased in terms of Fourier components as:
\be
\label{q}
p(x) \in \Pp_n^k \ \ \mbox{iff}
\ \
\left\{
\ba{l}
q(y)=0 \ \ {\rm if} \ \ 1\le \Wt(y)\le k,\\
q(0,\ldots,0)=1,\\
\mbox{for all} \ x\in \Bb^n \\
\sum_{y\in \Bb^n}
(-1)^{x\cdot y} q(y)\ge 0.\\
\ea
\right.
\ee
Note also that $\Pp_n^k$ is a convex set: if $p'(x),
p''(x)\in \Pp_n^k$ then for any $\alpha\in [0,1]$ we have $\alpha
p'(x) + (1-\alpha)p''(x) \in \Pp_n^k$.
 It is
known that the entropy is a concave functional, i.e.
$$
S[\alpha p'(x) + (1-\alpha) p''(x)] \ge \alpha S[p'(x)] +
(1-\alpha) S[p''(x)].
$$
It means that the minimum of $S[p(x)]$ over all $p(x)\in \Pp_n^k$
is achieved when $p(x)$ is an extremal point of $\Pp_n^k$. Because
$\Pp_n^k$ is specified by a finite number of linear equalities and
inequalities, it has a finite number of extremal
 points.
In principle we could find all them, compute $S[p(x)]$ for each
point and choose
 the
minimal value. However the set $\Pp_6^3$, which we are
interested in, has too many extremal points and such method doesn't
work in practice.

Instead we will proceed as follows. Consider six bit probability
distribution $p(x)$ defined by~(\ref{best}). An explicit
verification shows that $p(x) \in \Pp_6^3$ (as it should be,
because any measurement on the state $|\hex\ra$ produces
probability distribution from $\Pp_6^3$) and that $S[p(x)]=4$. It
tells us that
\be
\label{le4}
\inf_{ p(x)\in \Pp_6^3} S[p(x)] \le 4.
\ee
Also note that
\be
\label{5le6}
\inf_{p(x)\in \Pp_5^3} S[p(x)] \le \inf_{ p(x)\in \Pp_6^3} S[p(x)].
\ee
Indeed, take any $p(x)\in \Pp_6^3$ and average out the sixth bit.
Then, by definition, the distribution of the bits $1,2,\ldots,5$ is
3-uniform: $\sum_{x_6} p(x) \in \Pp_5^3$. The entropy
of the partial distribution can not exceed the entropy of the
joint distribution, so that
$S[\sum_{x_6} p(x)]$ $\le S[p(x)]$, which implies~(\ref{5le6}).
Now if we will manage to prove that
\be
\label{inf5}
\inf_{ p(x)\in \Pp_5^3} S[p(x)]=4,
\ee
then the work is done because~(\ref{le4},\ref{5le6},\ref{inf5})
imply~(\ref{inf6}). The remaining part of the text is the proof of
equality~(\ref{inf5}).

The proof is based on extremal point analysis as it was suggested
above. Let us parameterize $p(x)\in \Pp_5^3$ using its Fourier
transform $q(y)$, see~(\ref{Fourier},\ref{q}). The nonzero
components of $q(y)$ are listed below:
\be
\ba{cc}
q(1,1,1,1,1)\equiv q, &
q(1,1,0,1,1)\equiv q_3,\\
q(0,1,1,1,1)\equiv q_1, &
q(1,1,1,0,1)\equiv q_4,\\
q(1,0,1,1,1)\equiv q_2, &
q(1,1,1,1,0)\equiv q_5.\\
\ea \ee
Then $p(x)$ can be written as:
\be
\label{p(x)}
p(x) = (1/32) \{ 1+ (-1)^{ \Wt(x) }\,  [
q + \sum_{i=1}^5 q_i (-1)^{x_i}\, ] \}.
\ee
The positivity constraint $p(x)\ge 0$, $x\in \Bb^5$ specifies the
convex set $\Pp_5^3$ in the space of $q,q_1,\ldots,q_5$. If
$p(x)$ is an extremal point of $\Pp_5^3$ then $p(y)=0$ for at least
one $y\in \Bb^5$. By the symmetry, we can assume that
$p(0,0,0,0,0)=0$. So
 to find all
extremal points of $\Pp_5^3$ one suffices to find all extremal
points of the convex set
\be
\tilde{\Pp}_5^3 = \{ p(x) \in \Pp_5^3 \ : \ p(0,0,0,0,0)=0 \}.
\ee
In terms of variables $q,q_1,\ldots, q_5$ the set
$\tilde{\Pp}_5^3$ is described by the following linear
constraints:
\be
\tilde{\Pp}_5^3 = \left\{
\ba{l}
q=-1-\sum_{i=1}^5 q_i,\\
\sum_{i=1}^5 q_i \ge -1,\\
\displaystyle
q_i + q_j \le 0, \ \ 1\le i<j \le 5, \\
\ea
\right.
\ee
(fortunately, it appears that only part of inequalities $p(x)\ge 0$
is independent). It is convenient to introduce one more auxiliary
set $Q$ defined as
\be
Q = \{
(q_1,\ldots,q_5) \ : \ q_i + q_j \le 0, \ 1\le i < j \le 5 \}.
\ee
It is also a convex set. One can easily show that $Q$
has only one extremal point $q_1=\cdots=q_5=0$ and two types of
one-dimensional edges with five edges of each type coming out from
this extremal point:
\be
\label{edges}
\ba{rclr}
e^{(1)}_i &=& \{ q_i \le 0, \ q_j=0 \ \mbox{if}\ j\ne i \}, & i\in
[1,5], \\
e^{(2)}_i &=& \{ q_i \ge 0, \ q_j=-q_i, \mbox{if}\ j\ne i
\},  & i \in [1,5]. \\
\ea
\ee
The extremal points of $\tilde{\Pp}_5^3$ are those extremal points
of $Q$ for which $\sum_{i=1}^5 q_i \ge -1$ and also the
intersections of one-dimensional edges of
 $Q$
with the hyperplane $\sum_{i=1}^5 q_i =-1$. Summarizing, there are
only eleven extremal points of $\tilde{\Pp}_5^3$:\\
1) $q=-1$,
$q_1=\ldots=q_5=0$, \\
2) $q=0$, $q_i =-1$, $q_j=0$ if $j\ne i$;
$i\in [1,5]$,\\
3) $q=0$, $q_i= 1/3$, $q_j=-1/3$ if $j\ne
i$; $i\in [1,5]$.\\
Here extremal points 2) and 3) represent the
intersections of $e^{(1)}_i$ and $e
^{(2)}_i$
correspondingly with the hyperplane $\sum_{i=1}^5 q_i =-1$ while 1)
is the extremal point of $Q$. Substituting them into~(\ref{p(x)})
one can find the corresponding distributions $p(x)$.
One can check that for extremal points 1) and 2) the probability
$p(x)$ takes only values $0$ and $1/16$ thus having entropy
$S[p(x)]=4$. For extremal points 3) the probability takes the
values $0$, $1/12$, and  $1/24$. The entropy appears to be
$S[p(x)]=17/6 +
\log_2 3 \approx 4.4$.
Thus for all extremal points of $\Pp_5^3$ we have $S[p(x)
]\ge 4$
and for some extremal points $S[p(x)]=4$
which implies the equality~(\ref{inf5}).


\begin{references}
\bibitem{NC} M. Nielsen and I. Chuang,
{\it Quantum Computation and Quantum
Information} \ (Cambridge Univ. Press, 2000).

\bibitem{Vi} G. Vidal, J. Mod. Opt. {\bf 47}, 355 (2000).

\bibitem{BBPS} C. Bennett, H. Bernstein,
S. Popescu, and B. Schumacher,
Phys. Rev. A {\bf 53}, 2046  (1996).

\bibitem{PR} S. Popescu and D. Rohrlich,
Phys. Rev.  A {\bf 56}, R3319  (1997).

\bibitem{DW} K. Dennison and W. Wooters, LANL e-print quant-ph/0106058.

\bibitem{CS} A. Calderbank, E. Rains,
P. Shor, and N. Sloane, LANL e-print quant-ph/9608006.

\bibitem{bin} Binary matrix $M$ is
non-degenerate if for any binary vector $x\ne \vec{0}$
we have  $Mx\ne \vec{0}$ over binary field.
Equivalently, $M$ is nondegenerate if
${\rm det}(M)=1\, ({\rm mod}\,2)$.

\bibitem{est} According to Section~\ref{Lower},
it tells us that $S[\hex]\ge 3$.
This estimate however do not take into account that
$\rho_w$ is absolutely uniform for {\it any} partition $V=B\cup W$.

\bibitem{remark} Clearly, the outcomes distribution has the same
property for all choices of the bases $\calB_i$, not only for the
optimal one.

\end{references}
\end{document}